\documentclass[12pt]{article}

\catcode`\@=11
\@addtoreset{equation}{section}

\global\arraycolsep=2pt
\usepackage{mathrsfs,amsbsy,amssymb,latexsym,amsfonts,amsmath,cite}
\usepackage{graphicx,color}

\allowdisplaybreaks

\begin{document}

\begin{flushright}
\parbox{4cm}
{KUNS-2523}
\end{flushright}

\vspace*{1.5cm}

\begin{center}
{\Large \bf Minimal surfaces
in  $q$-deformed AdS$_5 \times $S$^5$
\\ with  Poincare coordinates}
\vspace*{1.5cm}\\
{\large Takashi Kameyama\footnote{E-mail:~kame@gauge.scphys.kyoto-u.ac.jp} 
and Kentaroh Yoshida\footnote{E-mail:~kyoshida@gauge.scphys.kyoto-u.ac.jp}} 
\end{center}

\vspace*{0.25cm}

\begin{center}
{\it Department of Physics, Kyoto University, \\ 
Kyoto 606-8502, Japan} 
\end{center}

\vspace{2cm}

\begin{abstract}
We study minimal surfaces in
$q$-deformed AdS$_5 \times $S$^5$
with a new coordinate system introduced in the previous work
1408.2189. 
In this letter, we introduce Poincare coordinates for the deformed theory. 
Then we construct minimal surfaces whose boundary shape is a circle. 
The
solution corresponds to a 1/2 BPS circular Wilson loop in the $q\to 1$ limit. 
A remarkable point is that the classical Euclidean action is not divergent unlike the original one. 
This finiteness indicates that the $q$-deformation may be regarded as a UV regularization.
\end{abstract}

\setcounter{footnote}{0}
\setcounter{page}{0}
\thispagestyle{empty}

\newpage

\section{Introduction}

The AdS/CFT correspondence 
\cite{M} is a realization of the equivalence between string theories and gauge theories. 
The most well-studied example is the duality between type IIB string theory on AdS$_5\times$S$^5$ 
and $\mathcal{N}=4$ super Yang-Mills theory.
It is often likened to a ``harmonic oscillator" among the gauge/gravity dualities.
A great discovery in the recent is an integrable structure behind 
the
AdS/CFT 
\cite{AF,review}. The integrability plays an important role in testing the  
conjectured relations in 
the
AdS/CFT. In particular,  
type IIB superstring theory on AdS$_5\times$S$^5$ is realized 
as a supersymmetric coset sigma model \cite{MT}
and its classical integrability is discussed in \cite{BPR}.

\medskip

The integrable structure behind the
AdS/CFT opens up a new arena 
for studying integrable deformations as well.
It would be of significance to reveal a deeper integrable structure 
behind gauge/gravity dualities beyond the AdS/CFT.
To tackle this issue, it is desirable to employ systematic ways to 
consider integrable deformations of two-dimensional non-linear sigma models. 
One way is to follow the Yang-Baxter sigma model approach \cite{Klimcik}. 
It has recently attracted a great deal of attention. 
According to this approach, an integrable deformation is fixed by picking up 
a skew-symmetric classical $r$-matrix satisfying the 
modified classical Yang-Baxter equation (mCYBE).

\medskip

It was originally proposed for principal chiral models \cite{Klimcik} and 
generalized to symmetric cosets \cite{DMV}\footnote{
For earlier developments on $q$-deformations of $su(2)$ and $sl(2)$\,, 
see \cite{KYhybrid,KMY-QAA,KOY,Sch}.}. Based on this generalization, 
the classical action of a $q$-deformed AdS$_5\times$S$^5$ superstring 
was constructed in \cite{DMV-string}. 
Then the metric\footnote{We refer to  it as the ABF metric.
} and NS-NS two-form have been fixed in \cite{ABF},
although, the dilaton and R-R sector have not been determined yet. 
In particular, a singular surface exists in the $q$-deformed AdS$_5$\,. 
For this deformed string theory, many works have been done.  
A certain limit of the deformed AdS$_n\times$S$^n$ has been studied in \cite{HRT}.
A mirror description was proposed in \cite{mirror}. 
The fast-moving string limit and the corresponding spin chain were studied in \cite{KY-LL}.
Giant magnon solutions have been studied in \cite{mirror,Kluson,Ahn,Banerjee}.
Deformed Neumann models were also derived in \cite{AM-NR}.
A new coordinate system to study the holographic relation was introduced 
in \cite{Kame-coord}. For $q$-deformations at root of unity, see \cite{Hollowood,Daniel}. 

\medskip

There exists another kind of integrable deformations based on 
the classical Yang-Baxter equation (CYBE), rather than the mCYBE. 
This approach is closely related to Jordanian deformations of the AdS$_5 \times $S$^5$ superstring 
\cite{KMY-JordanianAdSxS}. Classical $r$-matrices found in \cite{KMY-SUGRA,MY1,MY2,CMY-T11} 
lead to type IIB supergravity solutions including $\beta$-deformed backgrounds 
\cite{LM} and gravity  duals for non-commutative gauge theories \cite{HI}.  
The correspondence of this type is referred as the gravity/CYBE correspondence 
(For a short summary, see \cite{CYBE}). 

\medskip

In this letter, we are interested in the $q$-deformed AdS$_5 \times $S$^5$ superstring \cite{DMV-string}. 
An interesting issue is to consider a holographic relation in the $q$-deformed background. 
A proposal is that the singularity surface in the ABF metric \cite{ABF} may be regarded 
as the holographic screen \cite{Kame-coord}.  
Thus, it is nice to introduce a new coordinate system which describes only the spacetime 
enclosed by the singularity surface \cite{Kame-coord}.
With this coordinate system, minimal surfaces whose boundary is a straight line 
have been considered in \cite{Kame-coord}. The solutions are reduced to the well-known results 
\cite{Wilson,Kawamoto} in the $q\to 1$ limit. It is quite suggestive that 
the relations of the conserved charges satisfy the standard results in the usual AdS/CFT 
even though the supersymmetry is also $q$-deformed. 

\medskip 

We will here continue to study minimal surfaces
in the Euclidean $q$-deformed AdS space
by introducing the Poincare coordinates. 
Then we construct minimal surfaces whose 
boundary shape is a circle. A solution corresponds to a 1/2 BPS circular Wilson loop 
in the undeformed limit. A remarkable point is that the classical Euclidean action 
is not divergent unlike the original one. 
The finiteness 
indicates that the $q$-deformation may be regarded as a UV regularization. 

\medskip

This letter is organized as follows. 
Section 2 explains a new coordinate system for the $q$-deformed AdS$_5 \times $S$^5$ background.
Then we introduce the associated Poincare coordinate system for the deformed AdS part. 
In section 3, we consider two types of minimal surfaces which end up with a circle at the boundary. 
The former is constructed by supposing that 
the surface extends only within the deformed AdS$_2$ subspace. 
The latter is a generalization of the former including a polar angle in the deformed S$^2$\,.
The solutions corresponds to a 1/2 BPS circular Wilson loop \cite{BCFM,DGO} 
and a 1/4 BPS one \cite{Drukker}, respectively,
in the undeformed limit $q\to 1$\,.
A remarkable point is that in both cases, the resulting classical action 
is finite even though there is no UV cut-off.
In section 4, we consider a minimal surface with a cylinder shape.
The resulting classical action is not divergent unlike the undeformed one.
Section 
5
is devoted to conclusion and discussion.
Appendix A argues space-like geodesics to the singularity surface  in the deformed AdS.
In Appendix 
B
, we introduce the Poincare coordinates for the ABF metric \cite{ABF}.

\section{A $q$-deformation of AdS$_5\times$S$^5$}

We consider the bosonic part of the classical action of a $q$-deformed 
AdS$_5\times$S$^5$ superstring \cite{DMV-string}.
With the conformal gauge, the bosonic action (in the string frame) 
is composed of the metric part and the Wess-Zumino (WZ) term.

\medskip

In the following, we are concerned with minimal surfaces ending at the singularity surface 
in the coordinate system introduced in \cite{ABF}. 
It is helpful to employ a new coordinate system 
which describes the spacetime only inside the singularity surface \cite{Kame-coord}.
In the new coordinates, the singularity surface is located at the boundary. 

\medskip 

Let us first introduce the metric part and the WZ term of the bosonic action 
with the new coordinate system \cite{Kame-coord}. Then we introduce 
the associated Poincare coordinates. 

\subsection{The bosonic part of the $q$-deformed action}

We consider the bosonic part of the classical action of 
the $q$-deformed AdS$_5\times$S$^5$ superstring. 

\medskip

The bosonic action (in the conformal gauge) is composed of the metric part $S_G$ 
and the Wess-Zumino (WZ) term $S_{\rm WZ}$ like 
\begin{eqnarray}	
S &=& S_G + S_{\rm WZ}\,. \nonumber 
\end{eqnarray} 
Here $S_{\rm WZ}$ describes the coupling of string to an NS-NS two-form.  

\medskip 

The metric part $S_G$ is divided into the deformed AdS and internal sphere parts;  
\begin{eqnarray}	
S_G &=& \int\!d\tau d\sigma\, \left[\,
\mathcal{L}^G_{\rm AdS} + \mathcal{L}^G_{\rm S}\,
\right] \nonumber \\
& = &-\frac{1}{4\pi \alpha'}\int\!\!d\tau d\sigma\,\eta_{\mu\nu}
\left[G^{MN}_{\rm AdS}\partial^{\mu}X_M \partial^{\nu}X_N +
G^{PQ}_{\rm S}\partial^{\mu}Y_P \partial^{\nu}Y_Q \right]\,, \nonumber 
\end{eqnarray}
where the string world-sheet coordinates are $\sigma^{\mu}=(\tau,\sigma)$ 
with $\eta_{\mu\nu}=(-1,+1)$\,. 

\medskip 

\begin{figure}[htbp]
\begin{center}
\includegraphics[bb=0 0 720 405,scale=.45]{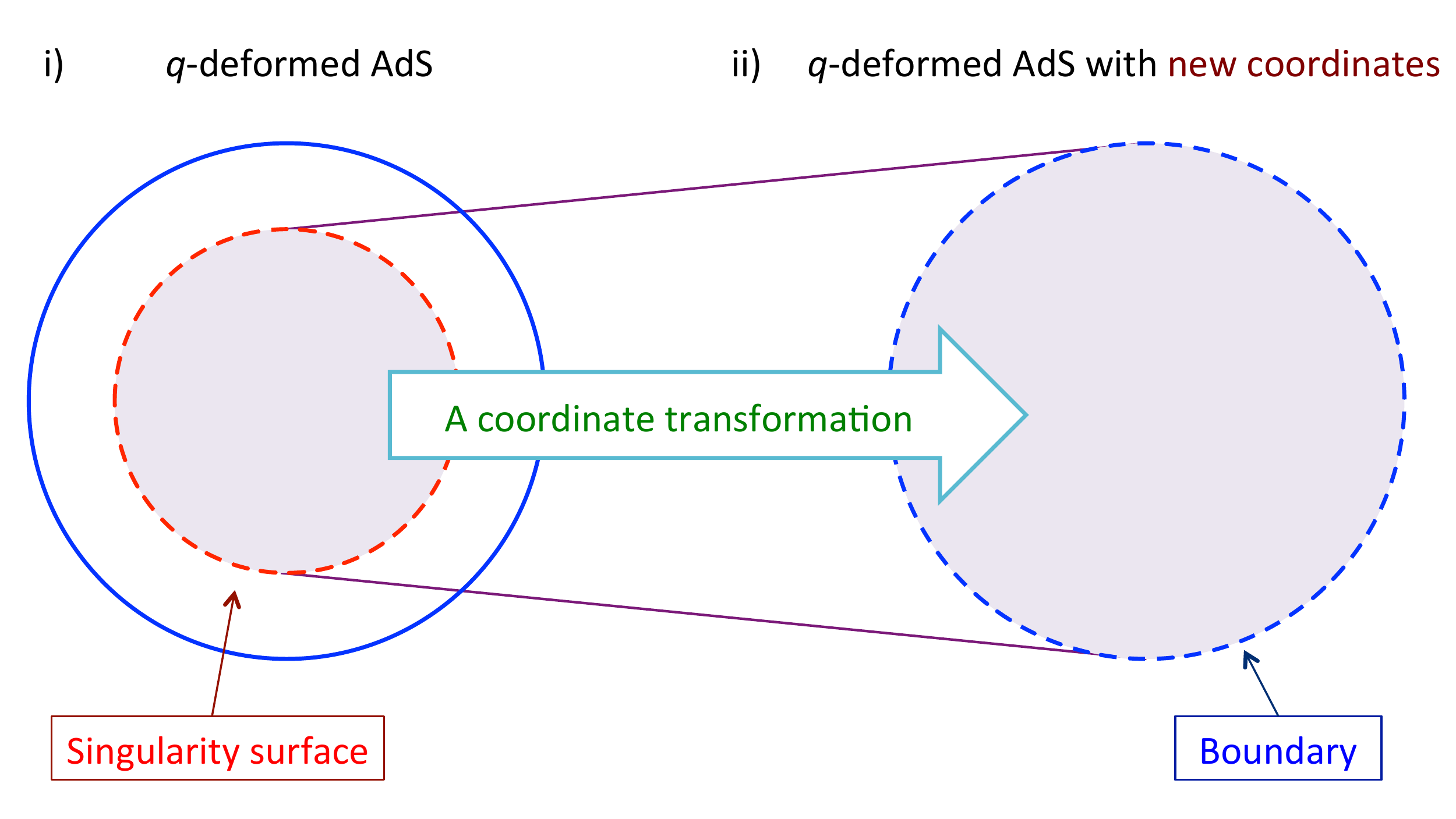}
\caption{\footnotesize A new coordinate system for the $q$-deformed AdS$_5$\,. 
The resulting geometry is 
enclosed by 
the singularity surface.
\label{1:fig}}
\end{center}
\end{figure}

With a new coordinate system proposed in \cite{Kame-coord}, 
the metric for the deformed AdS and sphere parts are given by, respectively,
\begin{eqnarray}
ds^2_{\textrm{AdS}_5} &=& R^2\sqrt{1+C^2}
\Bigl[ -\cosh^2\chi\,dt^2
+\dfrac{d\chi^2}{1+C^2\cosh^2\chi}\label{ads}\\ 
&&\hspace{-1cm}
+\dfrac{(1+C^2\cosh^2\chi)\sinh^2\chi}{(1+C^2\cosh^2\chi)^2
+C^2\sinh^4\chi\sin^2\zeta}\,(d\zeta^2+\cos^2\zeta\,d\psi_1^2)
+\dfrac{\sinh^2\chi\sin^2\zeta\,d\psi_2^2}{1+C^2\cosh^2\chi}\Bigr]\,,\nonumber
\end{eqnarray}
\begin{eqnarray}
ds^2_{\textrm{S}^5} &=& R^2\sqrt{1+C^2}
\Bigl[ \cos^2\gamma\,d\phi^2
+\dfrac{d\gamma^2}{1+C^2\cos^2\gamma}\label{s}\\ 
&&\hspace{.cm}
+\dfrac{(1+C^2\cos^2\gamma)\sin^2\gamma}{(1+C^2\cos^2\gamma)^2
+C^2\sin^4\gamma\sin^2\xi}\,(d\xi^2+\cos^2\xi\,d\phi_1^2)
+\dfrac{\sin^2\gamma\sin^2\xi\,d\phi_2^2}{1+C^2\cos^2\gamma}\Bigr]\,.
\nonumber
\end{eqnarray}
Here the deformed AdS$_5$ is parameterized by 
the coordinates $(t\,, \psi_1\,, \psi_2\,,\zeta\,, \chi)$\,, while
the deformed S$^5$ is described  by $(\phi\,, \phi_1\,, \phi_2 \,, \xi\,, \gamma)$\,.  
The deformation is characterized by a real parameter $C \in [0,\infty)$\,. 
When $C=0$\,, the geometry is reduced to the undeformed AdS$_5\times$S$^5$ 
with the curvature radius $R$\,. 
A relation between the ABF coordinates and the new ones is depicted in Fig.\,\ref{1:fig}.
It should be mentioned that  a curvature singularity exists 
at $\chi=\infty$ in the new coordinate system as well.
In \cite{Kame-coord}, it is shown that
the proper 
time
from any point to the singularity is infinite, 
while it does not take infinite time to reach the singularity in the coordinate time. 
This is the same result as in the usual AdS space with the global coordinates. 
Thus it seems likely to regard the singularity surface 
as the boundary in the holographic setup for the $q$-deformed geometry.

\medskip 

The WZ term $S_{\rm WZ}$ 
 for the AdS part and the sphere part are given by, respectively,  
\begin{eqnarray}
	\mathcal{L}_{\textrm{AdS}}^{\rm WZ} &=& \frac{T_{(C)}}{2}\,C\,
\epsilon^{\mu\nu}\dfrac{\sinh^4\chi\sin2\zeta}{(1+C^2\cosh^2\chi)^2
+C^2\sinh^4\chi\sin^2\zeta}\,\partial_\mu\psi_1\partial_\nu\zeta\,, 
\label{wz_ads_new} \\  
\mathcal{L}_{\textrm{S}}^{\rm WZ} &=& -\frac{T_{(C)}}{2}\,C\,
\epsilon^{\mu\nu}\dfrac{\sin^4\gamma\sin2\xi}{(1+C^2\cos^2\gamma)^2
+C^2\sin^4\gamma\sin^2\xi}\,\partial_\mu\phi_1\partial_\nu\xi\,.
\label{wz_s_new}
\end{eqnarray}
Here the totally anti-symmetric tensor $\epsilon^{\mu\nu}$ is normalized as $\epsilon^{01}=+1$\,. 
We have introduced a $C$-dependent string tension $T_{(C)}$ defined as
\begin{eqnarray}
T_{(C)}\equiv T\sqrt{1+C^2}\,, \qquad 
T\equiv \dfrac{R^{2}}{2\pi\alpha'}\,, \qquad 
\sqrt{\lambda}\equiv\dfrac{R^{2}}{\alpha'}\,.
\end{eqnarray}
Each component of the WZ term is proportional to $C$\,, 
and hence it vanishes when $C=0$\,.

\subsection{Poincare coordinates}

So far, we have discussed in the global coordinates. However, for our purpose, 
it is desirable to introduce the Poincare coordinates for the deformed AdS part (\ref{ads}).  

\medskip 

Let us first perform the following coordinate transformation
\begin{eqnarray}
\cosh\chi\equiv\frac{1}{\cos\theta}\,,
\end{eqnarray}
and the Wick rotation $t \to i\tau$ 
to move to the Poincare coordinates
. 
The  metric (\ref{ads}) is rewritten as the 
deformed 
Euclidean
  AdS$_5$\,,
\begin{eqnarray}
ds^{2}_{\textrm{AdS}_5}&=& R^2\sqrt{1+C^2}
\Bigl[ \frac{d\tau^2}{\cos^2\theta}
+\dfrac{d\theta^2}{\cos^2\theta+C^2} \nonumber\\ 
&&\hspace{-1cm}+\dfrac{(\cos^2\theta+C^2)\sin^2\theta}{(
\cos^2\theta+C^2)^2+C^2\sin^4\theta\sin^2\zeta}\,(d\zeta^2+\cos^2\zeta\,d\psi_1^2 )
+\frac{\sin^2\theta\sin^2\zeta\,d\psi_2^2}{\cos^2\theta+C^2}\Bigr]\,.\label{ads'}
\end{eqnarray}
Then, performing a coordinate transformation\,:
\begin{eqnarray}
z\equiv \textrm{e}^\tau\cos\theta\,,\qquad r\equiv\textrm{e}^\tau\sin\theta\,, 
\end{eqnarray}
we can obtain the following metric, 
\begin{eqnarray}
ds^{2}_{\textrm{AdS}_5}&=& R^2\sqrt{1+C^2}
\biggl[ \frac{dz^2+dr^2}{z^2+C^2(z^2+r^2)}
+ \frac{C^2(z\,dz+r\,dr)^2}{z^2\bigl(z^2+C^2(z^2+r^2)\bigr)}
\nonumber\\ 
&&\hspace{-1cm}
+ \frac{\bigl(z^2+C^2(z^2+r^2)\bigr)r^2}{
\bigl(z^2+C^2(z^2+r^2)\bigr)^2+C^2r^4\sin^2\zeta}\,(d\zeta^2+\cos^2\zeta\,d\psi_1^2 ) 
+  \frac{r^2\sin^2\zeta\,d\psi_2^2}{z^2+C^2(z^2+r^2)}
\biggr]
\,.\label{poincare}
\end{eqnarray}
When $C=0$\,, one can reproduce the 
Euclidean 
 AdS$_5$ metric with the Poincare coordinates. 
In this sense, the metric (\ref{poincare}) may be regarded as 
the Poincare version of the deformed AdS$_5$\,. 
Note that the Lorentzian signature can usually be recovered 
via the relation 
$dr^2 + r^2 d\Omega_3^2 = dt_{\rm P}^2 + \sum_{i=1}^3dx_i^2$\,, 
where $t_{\rm P}$ is the (Euclidean) Poincare time,
but for the deformed AdS geometry above it seems difficult to move back to the Lorentzian signature directly. 

\medskip 

After the Wick rotation has been performed, all we are concerned with 
are space-like paths rather than time-like and null ones. 
As argued in App.\ A, the space-like proper distance to the singularity surface is finite 
when $C \neq 0$\,, in comparison to the undeformed case with $C=0$\,. 
This property will be crucial in the next section.

\section{Minimal surfaces}

Let us study minimal surfaces in the $q$-deformed AdS$_5\times$S$^5$ string 
with the Poincare coordinates.
Note that  we are working  in the Euclidean signature and also on the sting world-sheet.
In particular, we consider minimal surfaces whose boundary shape is a circle
for two cases, 1) the AdS$_2$ subspace 2) the AdS$_2\times$S$^2$ subspace.
In both cases, for non-vanishing $C$\,, 
the classical actions are finite without introducing a UV cut-off.  
In the $q\to 1$ limit, the solutions are reduced to minimal surfaces dual  
for a 1/2 BPS circular Wilson loop \cite{BCFM,DGO} and a 1/4 BPS circular Wilson loop 
\cite{Drukker}, respectively.

\subsection{A deformed AdS$_2$ subspace}

We shall consider minimal surfaces which ends up with a circle 
at the boundary of the $q$-deformed AdS$_5$ with the Poincare coordinates (\ref{poincare})\,.
Let us consider an ansatz with the conformal gauge\,:
\begin{eqnarray}
z =\sqrt{a^2-r^2}\,,\qquad r=r(\sigma)\,,\qquad\psi_1=\psi_1(\tau)\,,\qquad\psi_2=\zeta=0\,,
\label{circular-ansatz}
\end{eqnarray}
with $0\leq\tau<2\pi\,,\,0\leq\sigma<\infty$\,.
Here $a$ is the radius of the circle at the boundary. 
Note that (\ref{circular-ansatz}) is a consistent ansatz 
and the WZ term vanishes under (\ref{circular-ansatz}) as explained in \cite{HRT}\,. 
Then the geometry with the metric (\ref{poincare}) is reduced to the following deformed AdS$_2$\,,
\begin{eqnarray}
ds^{2}_{\textrm{AdS}_2}&=& \frac{R^2\sqrt{1+C^2}\,r^2}{(1+C^2)a^2-r^2}
\Bigl[ \frac{a^2\,dr^2}{r^2(a^2-r^2)}
+ d\psi_1^2
\Bigr]
\,.\label{ads2}
\end{eqnarray}
This metric leads to a solution of the string equation of motion\,:
\begin{eqnarray}
z =a\tanh \sigma\,,\qquad r=\frac{a}{\cosh \sigma}\,,\qquad \psi_1=\tau\,.
\label{circular}
\end{eqnarray}

\medskip

The next task is to evaluate the classical Euclidean action of this solution.
By following the undeformed case,
let us formally introduce a cut-off $\epsilon$ for the coordinate $z$\,.
It may be regarded as a cut-off $\sigma_0$ for the string world-sheet coordinate $\sigma$ 
through the classical solution (\ref{circular}) for $z$\,, 
\begin{eqnarray}
\epsilon = a\tanh \sigma_0\,.
\end{eqnarray}
Then the classical action is evaluated as
\begin{eqnarray}
S&=&\frac{\sqrt{\lambda}}{4\pi}\sqrt{1+C^2}\int_0^{2\pi}\!\!d\tau
\int_{\sigma_0}^\infty\!\!d\sigma\Bigl[\frac{2}{\sinh^2\sigma+C^2\cosh^2\sigma} \Bigr]\nonumber\\
&=&\sqrt{\lambda}\frac{\sqrt{1+C^2}}{C}\Bigl(\textrm{arccot}[C] 
-\textrm{arccot}\Bigl[\frac{Ca}{\epsilon}\Bigr]\Bigr)\,,
\label{S}
\end{eqnarray}
where $\lambda \equiv R^4/\alpha'^2$\,.
Note that the second term in (\ref{S}) can be expanded in terms of $\epsilon$ like 
\begin{eqnarray}
-\sqrt{\lambda}\frac{\sqrt{1+C^2}}{C}\,\textrm{arccot}\Bigl[\frac{Ca}{\epsilon}\Bigr]
=-\sqrt{\lambda}\frac{\sqrt{1+C^2}}{C}\,\Bigl(\frac{\epsilon}{Ca}+\mathcal{O}(\epsilon^2)\Bigr)\,,
\end{eqnarray}
where $C$ has been fixed in this expansion. 
One can readily see that the cut-off can be removed for non-vanishing $C$\,.
By taking $\epsilon\rightarrow0$ 
the classical action (\ref{S}) becomes\,,
\begin{eqnarray}
S=\sqrt{\lambda}\frac{\sqrt{1+C^2}}{C}\,\textrm{arccot}[C]\,.
\label{Sc}
\end{eqnarray}
It should be remarked 
that the action (\ref{Sc}) is
finite 
even if there is no UV cut-off 
for the string world-sheet (
equivalently for
the radial direction of the deformed AdS).
The result would come from the finiteness of the space-like proper distance to the singularity surface 
(See App.\ A). Then the deformation parameter $C$ works as a UV regularization 
and one does not need to introduce $\epsilon$ any more in evaluating the classical action. 
This point becomes clear by taking the $C\to 0$ limit of (
\ref{Sc}
).
By expanding  (
\ref{Sc}
) in terms of $C$\,, 
we obtain the following expression: 
\begin{eqnarray}
S=-\sqrt{\lambda}+\sqrt{\lambda}\,\frac{\pi}{2C}+\mathcal{O}(C)\,.
\label{Sc0}
\end{eqnarray}
However we need to include $\epsilon$ so as to reproduce the regularized result 
in the undeformed limit \cite{BCFM,DGO} as shown below. 
For this purpose, it is helpful to consider the undeformed limit of the classical action (\ref{S}) by taking $C\rightarrow 0$ while keeping $\epsilon$ finite. This corresponds to keeping the boundary of the solution 
away from the singularity surface. Then 
the result in the undeformed case \cite{BCFM,DGO} is reproduced, 
\begin{eqnarray}
S=-\sqrt{\lambda}+\sqrt{\lambda}\,\frac{a}{\epsilon}\,.
\label{c0}
\end{eqnarray}
Note that the divergent term is also reproduced in the undeformed limit 
and it can be canceled 
by taking account of a Legendre transformation as usual.

\medskip 

Note also that the action (\ref{Sc}) is independent of $a$ due to the scale invariance of the metric 
(\ref{poincare}) under $z \to c_0\,z$ and $r \to c_0\,r$\, where $c_0$ is a positive real constant.

\medskip 

It may be nice to write the classical action (\ref{Sc}) with the $q$-deformation parameter. 
The
deformation parameter $C$ is related to 
the $q$-deformation parameter through the following relation \cite{DMV-string,ABF},
\begin{eqnarray}
q=\textrm{e}^{-\nu/T}\,,\qquad\nu=\frac{C}{\sqrt{1+C^2}}\,.  
\label{q}
\end{eqnarray}
Then the result (\ref{Sc}) is expressed as
\begin{eqnarray}
S=\frac{\sqrt{\lambda}}{\nu}\,\textrm{arccot}\Bigl[\frac{\nu}{\sqrt{1-\nu^2}}\Bigr]\,.
\end{eqnarray}

\subsubsection*{The Legendre transformation}

It would be worth mentioning about an additional contribution coming from the boundary \cite{DGO}.
In the undeformed case, it is well recognized that the classical action has a linear divergence 
and then it can be removed by considering a Legendre transformation. 
The origin of this additional contribution is the surface term which appears in taking a variation of the classical action 
to obtain the equations motion. This is just because the minimal surface has the boundary.  
Hence one needs to discuss this contribution in the deformed case as well.
By taking account of a Legendre transformation, 
i.e., adding a total derivative to the Lagrangian,
the total action is written as
\begin{eqnarray}
\tilde{S}&=&S+S_L\,,\qquad
S_L=
\int_0^{2\pi }\!\!d\tau\int_{\sigma_0}^\infty\!\!d\sigma\,\partial_\sigma\Bigl(z\,\frac{\partial L}{\partial(\partial_\sigma z)}\Bigr)\,.
\end{eqnarray}
The term $S_L$ is a total derivative
and it is evaluated as
\begin{eqnarray}
S_L&=&-\int_0^{2\pi }\!\!d\tau\,z\,\frac{\partial L}{\partial(\partial_\sigma z)}\biggl|_{\sigma=\sigma_0}
=-\sqrt{\lambda}\,\frac{\sqrt{1+C^2}\,\tanh\sigma_0}{\sinh^2\sigma_0+C^2\cosh^2\sigma_0}\nonumber\\
&=&-\sqrt{\lambda}\,\sqrt{1+C^2}\frac{\epsilon\,(a^2-\epsilon^2)}{a(\epsilon^2+C^2a^2)}\,.
\label{SL}
\end{eqnarray}
A remarkable point is that (\ref{SL}) vanishes in the limit $\epsilon \to 0$ when $C \neq 0$\,, 
and it does not contribute to the final expression of the action.

\medskip

The next is to consider the undeformed limit of $S_L$ 
so as to cancel the divergent term in the undeformed limit of $S$ (\ref{S})\,.
The term $S_L$ in (\ref{SL}) vanishes when $C \neq 0$\,, 
hence this is the first place one need to introduce  $\epsilon$ so as to 
keep the boundary away from the singularity surface. 
By taking the $C \to 0$ limit with $\epsilon$ fixed
, the resulting expression of $S_{L}$ 
is given by  
\begin{eqnarray}
S_L=-\sqrt{\lambda}\,\frac{a}{\epsilon}+\mathcal{O}(\epsilon)\,,
\end{eqnarray}
and it cancels the divergent term in (\ref{c0}) as usual.

\medskip

It would be helpful to comment on the cut-off $\epsilon$ and the deformation parameter $C$\,. 
When $C\neq0\,,$ there is no strict need to introduce $\epsilon$ 
in evaluating the classical action $S$ and the Legendre term $S_L$ vanishes.
However, if the limit $\epsilon \to 0$ is taken first,  or $\epsilon$ is not turned on, 
the finite result cannot be reproduced correctly in the $C \to 0$ limit.
In this scene, there is a subtlety of the order the two limits: 1) $\epsilon \to 0$ 
and 2) $C \to 0$\,. At least so far, a possible resolution is to take the limit $C \to 0$ first 
while keeping $\epsilon$ finite. In the following, this order of the limits is supposed.

\subsubsection*{A multiple string case}

Finally, 
let us 
comment on a generalization of the solution (\ref{circular}) 
with wrapping $k$ times around the circle.
The solution is described by
\begin{eqnarray}
z =a\tanh k\sigma\,,\qquad r=\frac{a}{\cosh k\sigma}\,,\qquad \psi_1=k\tau\,.
\end{eqnarray} 
By introducing a cut-off $\epsilon$ on $z$ through
\begin{eqnarray}
\epsilon = a\tanh k \sigma_0\,,
\end{eqnarray}
then the classical action can be evaluated as
\begin{eqnarray}
S=
\sqrt{\lambda}\frac{\sqrt{1+C^2}}{C}k\Bigl(\textrm{arccot}[C] 
-\textrm{arccot}\Bigl[\frac{Ca}{\epsilon}\Bigr]\Bigr)
\,. 
\label{S-k}
\end{eqnarray}
The second term in (\ref{S-k}) vanishes in the $\epsilon\rightarrow0$ limit 
with non-vanishing $C$\,.
Hence, when $C\neq0$\,, no UV cut-off has to be introduced as in the previous case.
In the $C\rightarrow 0$ limit with non-zero $\epsilon$, 
the classical action (\ref{S-k}) reproduces the undeformed result \cite{Drukker},
\begin{eqnarray}
S=-\sqrt{\lambda}\,k+\sqrt{\lambda}\,\frac{ak}{\epsilon}\,. \label{3.17}
\end{eqnarray}
As for the contribution of the Legendre transformation, when $C \neq 0$\,, 
it vanishes in the $\epsilon \to 0$ limit as well.  
In the undeformed limit $C \to 0$ with $\epsilon$ fixed, it cancels the divergent term in (\ref{3.17}).

\subsection{A deformed AdS$_2\times $S$^{2}$ subspace}

The next issue is to consider a generalization of the solution constructed 
in the previous subsection. As we will show below, the solution is 
reduced to a string solution dual to a 1/4 BPS circular Wilson loop \cite{Drukker} 
in the undeformed limit. 

\medskip 

Firstly, let us suppose the following ansatz with the conformal gauge\,:
\begin{eqnarray}
&&z =\sqrt{a^2-r^2}\,,\qquad r=r(\sigma)\,,\qquad\psi_1=\psi(\tau)\,,\qquad\psi_2=\zeta=0\,,\nonumber\\
&&\gamma=\gamma(\sigma)\,,\qquad\phi_1=\phi(\tau)
\,,\qquad\phi=\phi_2=\xi=0\,,
\label{ANSATZ}
\end{eqnarray}
with $0\leq\tau<2\pi$ and $0\leq\sigma<\infty$\,. 
Here $a$ is the radius of the circle at the boundary.
Note that (\ref{ANSATZ}) is a consistent ansatz 
and the WZ term vanishes.
Then the deformed AdS$_5 \times $S$^5$ is reduced to a deformed AdS$_2\times$S$^2$ 
with the metric 
\begin{eqnarray}
ds^{2}_{\textrm{AdS}_2\times S^{2}}={R^2\sqrt{1+C^2}}
\biggl[
\frac{r^2}{(1+C^2)a^2-r^2}
\Bigl[ \frac{a^2\,dr^2}{r^2(a^2-r^2)}
+ d\psi_1^2\Bigr]
+\frac{d\gamma^2+\sin^2\gamma\,d\phi_1^2}{1+C^2\cos^2\gamma}
\biggr]
\,.\label{ds-1/4}\nonumber
\end{eqnarray}

\medskip 

Here let us introduce a latitude angle $\theta_0$ for the sphere part, and the  boundary condition is set to
\begin{eqnarray}
\gamma(\sigma=0)=\frac{\pi}{2}-\theta_0\,.
\end{eqnarray}
When $\theta_0=\pi/2$\,, the surface in the sphere part is reduced to a point (the north pole).  
The resulting solution is just the deformed AdS$_2$ one discussed in section 3.1.

\medskip

Then the classical solution is given by
\begin{eqnarray}
&& z =a\tanh \sigma\,,\qquad r=\frac{a}{\cosh \sigma}\,,
\qquad  \sin\gamma=\frac{1}{\cosh(\varsigma_0\pm\sigma)} \,,\nonumber \\ 
&& \psi_1 =\tau\,, \qquad \phi_1=\tau\,.
\label{sol-1/4}
\end{eqnarray}
Here the signatures $\pm$ in $\sin\gamma$ correspond to 
the surface which  extends over the northern  (the short side) or southern  hemisphere.
An integration constant $\varsigma_0$ can be fixed by the boundary condition at $\sigma=0$\,, 
\begin{eqnarray}
\sin\gamma(\sigma=0)=\frac{1}{\cosh\varsigma_0}=\cos\theta_0\,.
\end{eqnarray}

\medskip

The remaining task is to evaluate the classical Euclidean action. 
Let us formally introduce a cut-off $\epsilon$ for  $z$\,. 
Then it can be converted into a cut-off for $\sigma$ at $\sigma_0$ 
through the classical solution (\ref{sol-1/4}) for $z$\,,
\begin{eqnarray}
\epsilon = a\tanh \sigma_0\,. 
\end{eqnarray}
Then the classical action is given by 
\begin{eqnarray}
S&=&\frac{\sqrt{\lambda}}{4\pi}\sqrt{1+C^2}\int_0^{2\pi}\!\!d\tau\int_{\sigma_0}^\infty\!\!
d\sigma\Bigl[\frac{2}{\sinh^2\sigma+C^2\cosh^2\sigma}
+ \frac{2}{\cosh^2(\varsigma_0\pm\sigma)+C^2\sinh^2(\varsigma_0\pm\sigma)}\Bigr]\nonumber\\
&=&\sqrt{\lambda}\frac{\sqrt{1+C^2}}{C}\biggl(\textrm{arccot}[C] 
-\textrm{arccot}\Bigl[\frac{Ca}{\epsilon}\Bigr]\nonumber\\
&&\hspace{2cm}+\textrm{arctan}[C]\mp\textrm{arctan}\!
\left[C\tanh\left(\textrm{arccosh}[\sec\theta_0]
\pm\textrm{arctanh}\!\left[\frac{\epsilon}{a}\right]
\right)\right]\biggr)\,.
\label{S-1/4}
\end{eqnarray}
The action (\ref{S-1/4}) can be expanded in terms of $\epsilon$ 
and there is no divergence with non-vanishing $C$\,. 
By taking the $\epsilon\rightarrow0$ limit ($C$\,:\,fixed), the action (\ref{S-1/4}) becomes
\begin{eqnarray}
S=\sqrt{\lambda}\frac{\sqrt{1+C^2}}{C}\Bigl(\textrm{arccot}[C] 
+\textrm{arctan}[C]\mp\textrm{arctan}[C\sin\theta_0]\Bigr)\,.
\label{S-1/4c}
\end{eqnarray}
Thus the expression (\ref{S-1/4c}) is finite
even if there is no UV cut-off for the string world-sheet.
It indicates again that the parameter $C$ plays a role as a UV regularization.
Note that $C$ can be converted into the $q$-deformation parameter through the relation (\ref{q}). 
Then the result (\ref{S-1/4c}) is expressed as
\begin{eqnarray}
S=\frac{\sqrt{\lambda}}{\nu}\Bigl(\textrm{arccot}\Bigl[\frac{\nu}{\sqrt{1-\nu^2}}\Bigr] 
+\textrm{arctan}\Bigl[\frac{\nu}{\sqrt{1-\nu^2}}\Bigr] 
\mp\textrm{arctan}\Bigl[\frac{\nu}{\sqrt{1-\nu^2}}\sin\theta_0\Bigr] \Bigr)\,.
\end{eqnarray}

\medskip 

It is worth seeing the undeformed limit. 
By taking the $C\rightarrow0$ limit, the classical action (\ref{S-1/4}) is reduced to
\begin{eqnarray}
S=\sqrt{\lambda}\,\left(\frac{a}{\epsilon}\mp\sin\theta_0\right)
+\mathcal{O}({\epsilon})
\,.
\end{eqnarray}
After subtracting the divergent term, 
the undeformed result \cite{Drukker} is precisely reproduced.

\medskip

As for the boundary contribution, there is no contribution of $S_L$ because it vanishes 
in the $\epsilon \to 0$ limit when $C\neq 0$\,, as in Sec.\ 3.1. 
Similarly, $S_L$  cancels the divergent term in the $C \to 0$ limit with $\epsilon$ fixed.

\subsubsection*{A multiple string case}

Finally, let us comment on a generalization of the solution (\ref{sol-1/4}) 
with wrapping $k$ and $m$ times around the circle in the deformed AdS$_2$ and the deformed S$^2$, respectively.
The classical solution is given by
\begin{eqnarray}
&& z =a\tanh k \sigma\,,\qquad r=\frac{a}{\cosh k\sigma}\,,
\qquad  \sin\gamma=\frac{1}{\cosh m(\varsigma_0\pm\sigma)} \,,\nonumber \\ 
&& \psi_1 =k\tau\,, \qquad \phi_1=m\tau\,.
\end{eqnarray}
The integration constant $\varsigma_0$ is fixed again by the boundary condition at $\sigma=0$\,, 
\begin{eqnarray}
\sin\gamma(\sigma=0)=\frac{1}{\cosh m\varsigma_0}=\cos\theta_0\,.
\end{eqnarray}

\medskip

By formally introducing a cut-off $\epsilon$ on $z$ through 
\begin{eqnarray}
\epsilon =a\tanh k\sigma_0\,,
\end{eqnarray}
then the classical action can be evaluated as
\begin{eqnarray}
S&=&\frac{\sqrt{\lambda}}{4\pi}\sqrt{1+C^2}\int_0^{2\pi}\!\!d\tau\int_{\sigma_0}^\infty\!\!
d\sigma\Bigl[\frac{2}{\sinh^2k\sigma+C^2\cosh^2k\sigma}\nonumber\\
&&\hspace{5cm}
+ \frac{2}{\cosh^2m(\varsigma_0\pm\sigma)+C^2\sinh^2m(\varsigma_0\pm\sigma)}\Bigr]\nonumber\\
&=&\sqrt{\lambda}\frac{\sqrt{1+C^2}}{C}\biggl[k\Bigl(\textrm{arccot}[C] 
-\textrm{arccot}\Bigl[\frac{Ca}{\epsilon}\Bigr]\Bigr)\nonumber\\
&&\hspace{0cm}+m\left(\textrm{arctan}[C]\mp\textrm{arctan}\!
\left[C\tanh \!\left(\textrm{arccosh}[\sec\theta_0]
\pm\frac{m}{k}\textrm{arctanh}\!\left[\frac{\epsilon}{a}\right]
\right)\right]\right)\biggr]\,.
\label{S-1/4k}
\end{eqnarray}
In the $\epsilon\rightarrow0$ limit with non-vanishing $C$\,,
the action (\ref{S-1/4k}) becomes
\begin{eqnarray}
S&=&\sqrt{\lambda}\frac{\sqrt{1+C^2}}{C}\biggl(k\,\textrm{arccot}[C] 
+m\Bigl(\textrm{arctan}[C]\mp\textrm{arctan}[C\sin\theta_0]
\bigr]\Bigr)\biggr)\,.
\end{eqnarray}
By taking the $C\rightarrow 0$ limit with non-zero $\epsilon$\,, 
the classical action (\ref{S-1/4k}) reproduces to the undeformed result \cite{Drukker},
\begin{eqnarray}
S=\sqrt{\lambda}\left(-k+k\frac{a}{\epsilon}+m\mp m\sin\theta_0\right)+\mathcal{O}({\epsilon})\,. \label{3.32}
\end{eqnarray}
As for the contribution of the Legendre transformation, when $C \neq 0$\,, 
it vanishes in the $\epsilon \to 0$ limit as well.  
In the undeformed limit $C \to 0$ with $\epsilon$ fixed, it cancels the divergent term in (\ref{3.32}).

\section{A minimal surface with a cylinder shape}

Due to the deformation, it seems difficult to construct a straight string solution 
with the Poincare coordinates, while it is possible to construct a cylinder shape solution.

\medskip 

Hence, in
this section, we shall consider a minimal surface solution with a cylinder shape. 
Then the solution ends with a circle on the boundary. 
So, with the conformal gauge, we are concerned with the following ansatz:
\begin{eqnarray}
z =z(\sigma)\,,\qquad r=a\,,\qquad\psi_1=\psi_1(\tau)\,,\qquad\psi_2=\zeta=0\,,
\label{cylinder-ansatz}
\end{eqnarray}
with $0\leq\tau<2\pi\,,\,0\leq\sigma<\infty$\,. Here $a$ is a positive real constant. 
Note that (\ref{cylinder-ansatz}) is a consistent ansatz and the WZ term vanishes.
Then the geometry associated with the solution is described by a deformed AdS$_2$ with the metric
\begin{eqnarray}
ds^{2}_{\textrm{AdS}_2}&=& \frac{R^2\sqrt{1+C^2}}{(1+C^2)z^2+C^2a^2}
\Bigl[ (1+C^2)\,dz^2
+ a^2\,d\psi_1^2
\Bigr]
\,.
\end{eqnarray}
This metric leads to a solution of the string equation of motion\,:
\begin{eqnarray}
z =\frac{a\,\sigma}{\sqrt{1+C^2}}\,,\qquad \psi_1=\tau\,.
\label{cylinder}
\end{eqnarray}

\medskip

The next task is to evaluate the classical Euclidean action for this solution.
Let us formally introduce a cut-off $\epsilon$ for  $z$\,.
Then it can be converted into a cut-off for $\sigma$ at $\sigma_0$ via the classical solution (\ref{cylinder}) for $z$\,,
\begin{eqnarray}
\epsilon = \frac{a\,\sigma_0}{\sqrt{1+C^2}}\,. 
\end{eqnarray}
Then the classical action is evaluated as
\begin{eqnarray}
S&=&\frac{\sqrt{\lambda}}{4\pi}\sqrt{1+C^2}\int_0^{2\pi}\!\!d\tau
\int_{\sigma_0}^\infty\!\!d\sigma\,\frac{2}{\sigma^2+C^2}\nonumber\\
&=&\sqrt{\lambda}\,\frac{\sqrt{1+C^2}}{C}\textrm{arctan}\Bigl[\frac{C}{\sqrt{1+C^2}}\,\frac{a}{\epsilon}\,\Bigr]\,.
\label{S-cylinder}
\end{eqnarray}
The action (\ref{S-cylinder}) can be expanded in terms of $\epsilon$ and there is no divergent term with non-zero $C$\,.
By taking $\epsilon\rightarrow0$\,,
the classical action (\ref{S-cylinder})  becomes\,,
\begin{eqnarray}
S=\sqrt{\lambda}\,\frac{\sqrt{1+C^2}}{C}\,\frac{\pi}{2}\,.
\label{Sc-cylinder}
\end{eqnarray}
The action (\ref{Sc-cylinder}) is 
finite and hence no UV cut-off is needed 
for the string world-sheet.
It is remarkable again that $C$ plays a role of UV regularization.
Note that $C$ can be converted into the $q$-deformation parameter via the relation (\ref{q}), 
and then (\ref{Sc-cylinder}) is expressed as
\begin{eqnarray}
S=\frac{\sqrt{\lambda}}{\nu}\,\frac{\pi}{2}\,.
\end{eqnarray}

\medskip

Let us comment on the undeformed limit of the classical action (\ref{S-cylinder}). 
By taking the $C\rightarrow 0$ limit ($\epsilon$\,:\,fixed), 
 the undeformed result mentioned in \cite{BCFM} is reproduced as 
\begin{eqnarray}
S=\sqrt{\lambda}\,\frac{a}{\epsilon}\,.
\label{Sc0-cylinder}
\end{eqnarray}
This divergent term is canceled out with the boundary contribution as we will see below.

\medskip

It is of importance to consider the boundary term via a Legendre transformation \cite{DGO}.
In the present case, the total derivative term is given by 
\begin{eqnarray}
S_L=-\int_0^{2\pi}\!\!d\tau\,z\,\frac{\partial L}{\partial(\partial_\sigma z)}\biggl|_{\sigma=\sigma_0}=-\sqrt{\lambda}\,\frac{(1+C^2)\,a\,\epsilon}{\epsilon^2+C^2(\epsilon^2+a^2)}\,.
\label{Sl}
\end{eqnarray}
When $C\neq 0$\,, $S_L$
vanishes in the $\epsilon \to 0$ limit and so it does not contribute to the final expression. 
On the other hand, in the undeformed limit $C\rightarrow 0$ limit ($\epsilon$\,:\,fixed)\,,  
$S_L$ is reduced to the following form, 
\begin{eqnarray}
S_L=-\sqrt{\lambda}\,\frac{a}{\epsilon}\,,
\end{eqnarray}
and it cancels the divergent term in (\ref{Sc0-cylinder})\,.

\section{Conclusion and discussion}

In this letter, we have discussed a $q$-deformation of the AdS$_5 \times $S$^5$ superstring.  
It is conjectured in \cite{Kame-coord} that a singularity surface in the ABF metric 
may be regarded as a holographic screen in the deformed theory. 
To look for some support for this conjecture, we have further considered  minimal surfaces 
whose boundary shape is a circle, by using the associated Poincare coordinates. 
A remarkable feature is that the classical Euclidean action is not divergent unlike the undeformed one. 
The finiteness comes from the fact that the $q$-deformation may be regarded as a UV regularization. 
As a matter of course, the divergent term is reproduced in the undeformed limit. 
That is, the undeformed limit becomes singular. 
This result may indicate that our conjecture would make sense. To gain more support, 
it would be interesting to consider other minimal surfaces like cusp Wilson loop 
by employing the Poincare coordinates.

\medskip

There exist many open problems. As a matter of course, 
the most interesting issue is to unveil the gauge-theory side dual to 
the $q$-deformation of the AdS$_5\times$S$^5$ superstring. 
To tackle it, the first thing we have to do is to understand 
the boundary geometry of the deformed AdS$_5$ with the Poincare coordinates. 
Naively, the boundary geometry degenerates.
Let us consider the metric (\ref{poincare}) 
and naively study the boundary geometry by following the usual prescription. 
For the finiteness of the boundary metric, the scale factor $\Omega^2 \equiv z^2(z^2+C^2(z^2+r^2))$ 
is multiplied to the metric (\ref{poincare}). Then, after taking the $z\to 0$ limit, 
the resulting metric is dimensionally reduced to $ds^2 = R^2\sqrt{1+C^2}\, C^2r^2dr^2$\,. This is not intrinsic to our scenario 
but a general problem which typically appears when considering non-asymptotically AdS spaces including 
the $q$-deformed AdS space. 
It may be interesting to try to resolve this issue by following \cite{Horava}, 
but we have no idea for its resolution so far.
Probably, it would be nice to consider an appropriate scaling limit of the deformed Poincare AdS$_5$\,. 
We hope we could report the result in the near future. 

\medskip 

To identify the dual gauge-theory side, our result may be a key ingredient. 
One plausible speculative scenario is the following. 
Firstly, deduce a candidate of the dual gauge-theory from the UV finiteness. 
Then one can compute the vacuum expectation value of a circular Wilson loop 
in the field-theory side by summing up rainbow diagrams \cite{ESZ} 
or using a localization technique \cite{Pestun}. 
As a result, the strong-coupling result can be extracted. 
If the result on the gauge-theory side agrees with 
the one we obtained in the string-theory side, we would get
 support for the candidate gauge theory.

\medskip

We believe that our results 
on minimal surfaces 
can play an important role 
in uncovering a possible gauge-theory dual.

\subsection*{Acknowledgments}

We are grateful to Shoichi Kawamoto, Takuya Matsumoto, Sotaro Sugishita 
and Stijn van Tongeren for useful discussions.
We also appreciate the referee for helpful comments, in regarding the finiteness of the action.
The work of TK was supported by the Japan Society for the Promotion of Science (JSPS). 
This work is supported in part by JSPS Japan-Hungary Research Cooperative Program.  

\section*{Appendix}

\appendix

\section{Proper distance to the singularity surface}
Here we consider space-like paths to the singularity surface in the global deformed AdS$_5$ (\ref{ads}) with the  Lorentzian signature and their  proper distances.

\medskip

Let us focus on a radial space-like geodesic $x=(t(t_A),\chi(t_A))$ which satisfies a unit speed parametrization,
\begin{eqnarray}
\dot{x}^2=-\cosh^2\chi\,\dot{t}^2+\frac{\dot{\chi}^2}{1+C^2\cosh^2\chi}=+1\,,
\end{eqnarray}
where $t_A$ is an affine parameter and $\dot{x}=dx/dt_A\,.$
For simplicity, we have omitted the overall factor $R^2\sqrt{1+C^2}$\,.
The path can be solved by extremizing the action,
\begin{eqnarray}
S=\int \!\!dt_A \,\left[-\cosh^2\chi\,\dot{t}^2+\frac{\dot{\chi}^2}{1+C^2\cosh^2\chi}\right]\,.
\end{eqnarray}
As a result, we obtain the following expressions: 
\begin{eqnarray}
E = \cosh^2\chi\,\dot{t}\,,\qquad 
\dot{\chi}^2=(1+C^2\cosh^2\chi)\,\frac{E^2+\cosh^2\chi}{\cosh^2\chi}\,.
\end{eqnarray}
Here $E$ is a conserved quantity (global AdS energy)
defined as $E\equiv -p_t$ with $p_t\equiv G_{tt}\,dt/dt_A$\,.

\medskip

Then the proper distance from the origin  to the singularity surface  
is given by
\begin{eqnarray}
t_A&=&\int_0^\infty\frac{d\chi}{\dot{\chi}}=\int_0^\infty\frac{d\chi\,\cosh\chi}{\sqrt{(1+C^2\cosh^2\chi)(E^2+\cosh^2\chi)}}\nonumber\\
&=&\frac{1}{\sqrt{1+C^2}}\,\textrm{K}\left[\frac{1-C^2E^2}{1+C^2}\right]\,.
\label{proper}
\end{eqnarray}
This result indicates that in the deformed case 
the proper distance to the singularity surface is finite for any $C$\,,
while the space-like distance in the usual AdS$_5$ is infinite.
The purely radial case is obtained by taking $E=0$ giving
\begin{eqnarray}
t_A=\frac{1}{\sqrt{1+C^2}}\,\textrm{K}\left[\frac{1}{1+C^2}\right]\,.
\end{eqnarray}
The causal physical quantities in the Lorentzian signature are irrelevant to this result. 
However, in the Euclidean signature, this behavior of the space-like proper distance 
is responsible and would explain that the minimal surface areas are finite even if 
no UV cut-off has been introduced.

\section{Poincare coordinates for the ABF metric}

We shall introduce the Poincare coordinates for the ABF metric \cite{ABF}. 

\medskip 

The ABF metric  for the AdS part is given by  
\begin{eqnarray}
ds^2_{\textrm{AdS}_5} &=& R^2(1+C^2)^{\frac{1}{2}}
\Bigl[ -\dfrac{\cosh^2\rho\,dt^2}{1-C^2\sinh^2\rho}
+\dfrac{d\rho^2}{1-C^2\sinh^2\rho}
+\dfrac{\sinh^2\rho\,d\zeta^2}{1+C^2\sinh^4\rho\sin^2\zeta}\nonumber\\ 
&&\hspace{2.0cm}+\dfrac{\sinh^2\rho\cos^2\zeta\,d\psi_1^2}{1+C^2\sinh^4\rho\sin^2\zeta}
+\sinh^2\rho\sin^2\zeta\,d\psi_2^2\Bigr]\,. \label{AdS5}
\end{eqnarray}
Here the deformed AdS$_5$ is parameterized by 
the coordinates $(t\,, \psi_1\,, \psi_2\,,\zeta\,, \rho)$\,.
The singularity surface exists at $\rho=\textrm{arcsinh}(1/C)$\,.
The deformation is measured by a real parameter $C \in [0,\infty)$\,. 

\medskip

Here we consider the Poincare coordinate for the deformed AdS$_5$ (\ref{AdS5}).
It is helpful to take a coordinate transformation
\begin{eqnarray}
\cosh\rho\equiv\frac{1}{\cos\vartheta}\,,
\end{eqnarray}
and perform the Wick rotation $t\to i\tau$\,.

\medskip

Then, performing a coordinate transformation 
\begin{eqnarray}
\textrm{z}\equiv \textrm{e}^\tau\cos\vartheta\,,\qquad 
\textrm{r}\equiv\textrm{e}^\tau\sin\vartheta\,,
\end{eqnarray}
the metric (\ref{AdS5}) can be rewritten into the  Poincare form, 
\begin{eqnarray}
ds^{2}_{\textrm{AdS}_5}&=& R^2\sqrt{1+C^2}
\Bigl[ \frac{d\textrm{z}^2+d\textrm{r}^2}{\textrm{z}^2-C^2\textrm{r}^2}
+ \frac{\textrm{r}^2(d\zeta^2+\cos^2\zeta\,d\psi_1^2)}{
\textrm{z}^2+C^2(\frac{\textrm{r}}{\textrm{z}})^2\textrm{r}^2\sin^2\zeta}+
\frac{\textrm{r}^2\sin^2\zeta\,d\psi_2^2}{\textrm{z}^2}
\Bigr]\,.
\label{ABF-poincare}
\end{eqnarray}
By taking the following coordinate transformation\,:
\begin{eqnarray}
\sqrt{1+C^2}\,z\equiv\sqrt{\textrm{z}^2-C^2\textrm{r}^2}\,,\qquad r\equiv\textrm{r}\,,
\end{eqnarray}
the resulting metric corresponds to the Poincare coordinates (\ref{poincare})\,.


\begin{thebibliography}{99}

\bibitem{M}  
  J.~M.~Maldacena,
  ``The large N limit of superconformal field theories and supergravity,''
  Int.\ J.\ Theor.\ Phys.\  {\bf 38} (1999) 1113
   [Adv.\ Theor.\ Math.\ Phys.\  {\bf 2} (1998) 231]
  [hep-th/9711200].

\bibitem{AF}
  G.~Arutyunov and S.~Frolov,
  ``Foundations of the AdS$_5\times$S$^5$ Superstring. Part I,''
  J.\ Phys.\ A {\bf 42} (2009) 254003
  [arXiv:0901.4937 [hep-th]].

\bibitem{review}
  N.~Beisert {\it et al.},
  ``Review of AdS/CFT Integrability: An Overview,''
  Lett.\ Math.\ Phys.\  {\bf 99} (2012) 3
  [arXiv:1012.3982 [hep-th]].

\bibitem{MT}
  R.~R.~Metsaev and A.~A.~Tseytlin,
  ``Type IIB superstring action in AdS$_5\times$S$^5$ background,''  
  Nucl.\ Phys.\ B {\bf 533} (1998) 109
  [hep-th/9805028].

\bibitem{BPR}
  I.~Bena, J.~Polchinski and R.~Roiban,
  ``Hidden symmetries of the AdS$_5\times$S$^5$ superstring,''
  Phys.\ Rev.\ D {\bf 69} (2004) 046002
  [hep-th/0305116].

\bibitem{Klimcik}
 C.~Klimcik,
  ``Yang-Baxter sigma models and dS/AdS T duality,'' JHEP {\bf 0212} (2002)
051 [hep-th/0210095];   ``On integrability of the Yang-Baxter sigma-model,'' J.\ Math.\
Phys.\ {\bf 50} (2009) 043508 [arXiv:0802.3518 [hep-th]]. 

\bibitem{DMV}
  F.~Delduc, M.~Magro and B.~Vicedo,
  ``On classical $q$-deformations of integrable $\sigma$-models,'' 
JHEP {\bf 1311} (2013) 192 [arXiv:1308.3581 [hep-th]].     
  
\bibitem{KYhybrid}
  I.~Kawaguchi and K.~Yoshida,
  ``Hybrid classical integrability in squashed sigma models,''
  Phys.\ Lett.\ B\ {\bf 705} (2011) 251
  [arXiv:1107.3662 [hep-th]]; 
   ``Hybrid classical integrable structure of squashed sigma models: A short summary,''  
  J.\ Phys.\ Conf.\ Ser.\  {\bf 343} (2012) 012055 
  [arXiv:1110.6748 [hep-th]]; 
  ``Hidden Yangian symmetry in sigma model on squashed sphere,''
  JHEP {\bf 1011} (2010) 032. 
  [arXiv:1008.0776 [hep-th]].     

\bibitem{KMY-QAA}
  I.~Kawaguchi, T.~Matsumoto and K.~Yoshida,
  ``The classical origin of quantum affine algebra in squashed sigma models,''  
  JHEP {\bf 1204} (2012) 115  [arXiv:1201.3058 [hep-th]];  
  ``On the classical equivalence of monodromy matrices in squashed sigma model,''  
  JHEP {\bf 1206} (2012) 082  [arXiv:1203.3400 [hep-th]].

\bibitem{KOY}
  I.~Kawaguchi, D.~Orlando and K.~Yoshida,
  ``Yangian symmetry in deformed WZNW models on squashed spheres,''
  Phys.\ Lett.\  B {\bf 701} (2011) 475. 
  [arXiv:1104.0738 [hep-th]]; I.~Kawaguchi and K.~Yoshida,
   ``A deformation of quantum affine algebra in squashed
 WZNW models,'' to appear in J.\ Math.\ Phys. [arXiv:1311.4696 [hep-th]].

\bibitem{Sch}
  I.~Kawaguchi and K.~Yoshida,
  ``Classical integrability of Schrodinger sigma models and $q$-deformed Poincare symmetry,''  
JHEP {\bf 1111} (2011) 094  [arXiv:1109.0872 [hep-th]]; 
  ``Exotic symmetry and monodromy equivalence in Schrodinger sigma models,''  
JHEP {\bf 1302} (2013) 024  [arXiv:1209.4147 [hep-th]]; 
  I.~Kawaguchi, T.~Matsumoto and K.~Yoshida,
  ``Schroedinger sigma models and Jordanian twists,''  
JHEP {\bf 1308} (2013) 013  [arXiv:1305.6556 [hep-th]].  

\bibitem{DMV-string} 
 F.~Delduc, M.~Magro and B.~Vicedo,
  ``An integrable deformation of the AdS$_5\times$S$^5$ superstring
action,'' Phys.\ Rev.\ Lett.\  {\bf 112} (2014) 051601
  [arXiv:1309.5850 [hep-th]];
 F.~Delduc, M.~Magro and B.~Vicedo,
  ``Derivation of the action and symmetries of the $q$-deformed AdS$_5 \times $S$^5$ superstring,''
  JHEP {\bf 1410} (2014) 132
  [arXiv:1406.6286 [hep-th]].

\bibitem{ABF} 
 G.~Arutyunov, R.~Borsato and S.~Frolov,
  ``S-matrix for strings on $\eta$-deformed AdS$_5\times$S$^5$,'' 
  JHEP {\bf 1404} (2014) 002
  [arXiv:1312.3542 [hep-th]].

\bibitem{HRT}
  B.~Hoare, R.~Roiban and A.~A.~Tseytlin,
  ``On deformations of AdS$_n\times$S$^n$ supercosets,''
  JHEP {\bf 1406} (2014) 002
  [arXiv:1403.5517 [hep-th]].

\bibitem{mirror}  
 G.~Arutynov, M.~de Leeuw and S.~J.~van Tongeren,
  ``On the exact spectrum and mirror duality of the (AdS$_5\times$S$^5$)$_{\eta}$ superstring,''
    arXiv:1403.6104 [hep-th];
  G.~Arutyunov and S.~J.~van Tongeren,
  ``The $\mathrm{AdS}_5 \times \mathrm{S}^5$ mirror model as a string,''
   Phys.\ Rev.\ Lett.\  {\bf 113} (2014) 261605
  [arXiv:1406.2304 [hep-th]].

\bibitem{KY-LL} 
  T.~Kameyama and K.~Yoshida,
  ``Anisotropic Landau-Lifshitz sigma models from $q$-deformed AdS$_5 \times $S$^5$ superstrings,''
  JHEP {\bf 1408} (2014) 110
  [arXiv:1405.4467 [hep-th]];
  ``String theories on warped AdS backgrounds and integrable deformations of spin chains,''
  JHEP {\bf 1305} (2013) 146
  [arXiv:1304.1286 [hep-th]].
  
\bibitem{Kluson} 
  M.~Khouchen and J.~Kluson,
  ``Giant Magnon on Deformed AdS$_3\times$S$^3$\,,''
    Phys.\ Rev.\ D {\bf 90} (2014) 066001
  [arXiv:1405.5017 [hep-th]].

\bibitem{Ahn} 
  C.~Ahn and P.~Bozhilov,
  ``Finite-size giant magnons on
$\eta$-deformed AdS$_5 \times $S$^5$,''
   Phys.\ Lett.\ B {\bf 737} (2014) 293
  [arXiv:1406.0628 [hep-th]].
  
\bibitem{Banerjee}  
  A.~Banerjee and K.~L.~Panigrahi, 
  ``On the Rotating and Oscillating strings in (AdS$_3\times $S$^3$)$_{\varkappa}$,''
  JHEP {\bf 1409} (2014) 048
  [arXiv:1406.3642 [hep-th]].
  
\bibitem{AM-NR} 
  G.~Arutyunov and D.~Medina-Rincon,
  ``Deformed Neumann model from spinning strings on (AdS$_5 \times $S$^5)_\eta$,''
  JHEP {\bf 1410} (2014) 50
  [arXiv:1406.2536 [hep-th]].

\bibitem{Kame-coord}
  T.~Kameyama and K.~Yoshida,
  ``A new coordinate system for $q$-deformed AdS$_5\times$S$^5$ and classical string solutions,''
  J.\ Phys.\ A {\bf 48} (2015) 7,  075401
  [arXiv:1408.2189 [hep-th]].

\bibitem{Hollowood}
  T.~J.~Hollowood, J.~L.~Miramontes and D.~M.~Schmidtt,
  ``Integrable Deformations of Strings on Symmetric Spaces,''
  JHEP {\bf 1411} (2014) 009
  [arXiv:1407.2840 [hep-th]];
  ``An Integrable Deformation of the AdS5 x S5 Superstring,''
  J.\ Phys.\ A {\bf 47} (2014) 49,  495402
  [arXiv:1409.1538 [hep-th]].

\bibitem{Daniel}
  K.~Sfetsos and D.~C.~Thompson,
  ``Spacetimes for $\lambda$-deformations,''
  JHEP {\bf 1412} (2014) 164
  [arXiv:1410.1886 [hep-th]].

\bibitem{KMY-JordanianAdSxS}
 I.~Kawaguchi, T.~Matsumoto and K.~Yoshida,
  ``Jordanian deformations of the AdS$_5\times$S$^5$ superstring,'' 
  JHEP {\bf 1404} (2014) 153
  [arXiv:1401.4855 [hep-th]]; 

\bibitem{KMY-SUGRA} 
 I.~Kawaguchi, T.~Matsumoto and K.~Yoshida,
  ``A Jordanian deformation of AdS space in type IIB supergravity,''
  JHEP {\bf 1406} (2014) 146
  [arXiv:1402.6147 [hep-th]]. 
  
\bibitem{MY1}  
T.~Matsumoto and K.~Yoshida,
  ``Lunin-Maldacena backgrounds from the classical Yang-Baxter equation 
-- Towards the gravity/CYBE correspondence,''
  JHEP {\bf 1406} (2014) 135
  [arXiv:1404.1838 [hep-th]]. 

\bibitem{MY2}
 T.~Matsumoto and K.~Yoshida,
  ``Integrability of classical strings dual for noncommutative gauge theories,''
  JHEP {\bf 1406} (2014) 163
  [arXiv:1404.3657 [hep-th]].

\bibitem{CMY-T11}
  P.~M.~Crichigno, T.~Matsumoto and K.~Yoshida,
  ``Deformations of $T^{1,1}$ as Yang-Baxter sigma models,''
  JHEP {\bf 1412} (2014) 085
  [arXiv:1406.2249 [hep-th]].

\bibitem{LM}
 O.~Lunin and J.~M.~Maldacena,
  ``Deforming field theories with $U(1) \times U(1)$ global symmetry and their gravity duals,''  
JHEP {\bf 0505} (2005) 033  [hep-th/0502086].

\bibitem{HI}
  A.~Hashimoto and N.~Itzhaki,
  ``Noncommutative Yang-Mills and the AdS / CFT correspondence,''
  Phys.\ Lett.\ B {\bf 465} (1999) 142
  [hep-th/9907166]. \\
  J.~M.~Maldacena and J.~G.~Russo,
  ``Large N limit of noncommutative gauge theories,''
  JHEP {\bf 9909} (1999) 025
  [hep-th/9908134].

\bibitem{CYBE}
  T.~Matsumoto and K.~Yoshida,
  ``Integrable deformations of the AdS$_5\times$S$^5$ superstring 
and the classical Yang-Baxter equation 
-- Towards the gravity/CYBE correspondence --,''
  J.\ Phys.\ Conf.\ Ser.\  {\bf 563} (2014) 1,  012020
  [arXiv:1410.0575 [hep-th]].

\bibitem{Wilson}
 S.~-J.~Rey and J.~-T.~Yee,
  ``Macroscopic strings as heavy quarks in large N gauge theory 
and anti-de Sitter supergravity,''
  Eur.\ Phys.\ J.\ C {\bf 22} (2001) 379
  [hep-th/9803001]; \\
J.~M.~Maldacena,
  ``Wilson loops in large N field theories,''
  Phys.\ Rev.\ Lett.\  {\bf 80} (1998) 4859
  [hep-th/9803002].
  
\bibitem{Kawamoto}
  N.~Drukker and S.~Kawamoto,
  ``Small deformations of supersymmetric Wilson loops and open spin-chains,''
  JHEP {\bf 0607} (2006) 024
  [hep-th/0604124].

\bibitem{BCFM}
  D.~E.~Berenstein, R.~Corrado, W.~Fischler and J.~M.~Maldacena,
  ``The Operator product expansion for Wilson loops and surfaces in the large N limit,''
  Phys.\ Rev.\ D {\bf 59} (1999) 105023
  [hep-th/9809188].

\bibitem{DGO}
  N.~Drukker, D.~J.~Gross and H.~Ooguri,
  ``Wilson loops and minimal surfaces,''
  Phys.\ Rev.\ D {\bf 60} (1999) 125006
  [hep-th/9904191].

\bibitem{Drukker}
  N.~Drukker and B.~Fiol,
  ``On the integrability of Wilson loops in $AdS_5 \times S^5$: Some periodic ansatze,''
  JHEP {\bf 0601} (2006) 056
  [hep-th/0506058];
  N.~Drukker,
  ``1/4 BPS circular loops, unstable world-sheet instantons and the matrix model,''
  JHEP {\bf 0609} (2006) 004
  [hep-th/0605151];
  N.~Drukker, S.~Giombi, R.~Ricci and D.~Trancanelli,
  ``Supersymmetric Wilson loops on $S^3$,''
  JHEP {\bf 0805} (2008) 017
  [arXiv:0711.3226 [hep-th]].
  
\bibitem{Horava}  
 P.~Horava and C.~M.~Melby-Thompson,
  ``Anisotropic Conformal Infinity,''
  Gen.\ Rel.\ Grav.\  {\bf 43} (2011) 1391
  [arXiv:0909.3841 [hep-th]].  
  
%
  
\bibitem{ESZ}
  J.~K.~Erickson, G.~W.~Semenoff and K.~Zarembo,
  ``Wilson loops in N=4 supersymmetric Yang-Mills theory,''
  Nucl.\ Phys.\ B {\bf 582} (2000) 155
  [hep-th/0003055].  
  
\bibitem{Pestun}
  V.~Pestun,
  ``Localization of gauge theory on a four-sphere and supersymmetric Wilson loops,''
  Commun.\ Math.\ Phys.\  {\bf 313} (2012) 71
  [arXiv:0712.2824 [hep-th]]. 

\end{thebibliography}
\end{document}